\documentclass[12pt]{iopart}

\usepackage{iopams}
\usepackage{hyperref}

\def\imagi{\mathrm{i}}
\newtheorem{theorem}{Theorem}

\newtheorem{lemma}{Lemma}
\newtheorem{corollary}{Corollary}

\begin{document}

\title[]{On the existence of effective potentials in time-dependent density functional theory}

\author{M Ruggenthaler$^1$, M Penz$^2$ and D Bauer$^{1,3}$}

\address{$^1$ Max-Planck-Institut f\"ur Kernphysik, Postfach 103980, 69029 Heidelberg, Germany}
\address{$^2$ Institute of Theoretical Physics, University of Innsbruck, Technikerstr. 25, 6020 Innsbruck, Austria}
\address{$^3$ Institut f\"ur Physik, Universit\"at Rostock, 18051 Rostock, Germany}
\ead{m.ruggenthaler@mpi-k.de}
\begin{abstract}
We investigate the existence and properties of effective potentials
in time-dependent density functional theory. We outline conditions
for a general solution of the corresponding Sturm-Liouville boundary
value problems. We define the set of potentials and
$v$-representable densities, give a proof of existence of the
effective potentials under certain restrictions, and show the set of
$v$-representable densities to be independent of the interaction.\\

This is an author-created, un-copyedited version of an article
accepted for publication in \textit{J. Phys. A: Math. Theor.}
\textbf{42} (2009) 425207. IOP Publishing Ltd is not responsible for
any errors or omissions in this version of the manuscript or any
version derived from it. The definitive publisher authenticated
version is available online at
\href{http://dx.doi.org/10.1088/1751-8113/42/42/425207}{doi:10.1088/1751-8113/42/42/425207}.
\end{abstract}


\vspace{2pc}

\section{Introduction}

The calculation of the wave-function of a fully interacting
many-body quantum system is a formidable challenge \cite{Kohn}.
Hence, feasible approaches to the quantum many-body problem are very
important if one is interested in the properties of complex
multi-particle systems. One such approach is Nobel-Prize-winning
density functional theory (DFT) \cite{Kohn, Dreizler_Gross}. The
main theorem of DFT is the Hohenberg-Kohn theorem
\cite{Hohenberg_Kohn}, which proves that the external potential of
an interacting $N$-body system uniquely defines the one-particle
density of the ground-state via minimization of the corresponding
energy functional. The mathematical foundations of DFT were
extensively investigated and a rigorous formulation is available
\cite{Lieb, Blanchard_Bruening}. Substantial extensions of the
ground-state theory can be found in the literature
\cite{Dreizler_Gross}, e.g. to excited states or to relativistic
systems.
\\
The mainstay of applications of the minimization principle of DFT is
due to Kohn and Sham \cite{Kohn_Sham}. The so-called Kohn-Sham
scheme uses an auxiliary system of noninteracting particles which
has the same energy and one-particle density as the interacting
system. To make contact to the physical system one introduces the
so-called exchange-correlation energy functional
\cite{Dreizler_Gross}. It accounts for the difference between the
combined kinetic and interaction energy of the interacting system
and the kinetic energy of the noninteracting Kohn-Sham system. The
variational minimization of the energy with respect to the density
leads to a set of coupled, nonlinear single-particle differential
equations, the so-called Kohn-Sham equations. It can be rigorously
proven that a self-consistent solution of these equations will
generate the exact one-particle ground-state density of the
corresponding interacting system \cite{Dreizler_Gross}. In practice,
however, the exchange-correlation energy functional is not known and
has to be approximated.
\\
An exact extension of DFT to time-dependent systems was given in
\cite{Runge_Gross} by Runge and Gross. By assuming the external
potentials to be Taylor expandable in time about $t=t_{0}$ they
could prove a one-to-one correspondence between time-dependent
densities and external potentials. However, the straightforward
extension of the Kohn-Sham scheme to the time-dependent case as
shown in \cite{Runge_Gross} led to the so-called symmetry-causality
paradox \cite{Gross95}. This flaw in the time-dependent extension of
DFT was soon realized to be connected to the naive application of
the usual variational principle of time-dependent quantum mechanics
to TDDFT \cite{vanLeeuwen01}. Only with an extension of the
Runge-Gross theorem by van Leeuwen in \cite{vanLeeuwen99} one was
able to justify a time-dependent Kohn-Sham scheme. There it was
shown that by successive solution of Sturm-Liouville boundary value
problems one can formally construct a unique effective potential
governing the time-evolution of the noninteracting system such that
it reproduces the interacting one-particle density. In view of this
theorem one does not need a variational approach analogously to
time-independent DFT and can obtain the exact one-particle density
via propagation of the time-dependent Kohn-Sham equations.
\\
Note that, although the extended Runge-Gross theorem
\cite{vanLeeuwen99} shows the uniqueness of the effective potential,
the conditions of existence were not investigated.
\\
\\
The intention of this work is not to give an introduction to TDDFT,
for this we refer to \cite{TDDFT}, but to consider the mathematical
foundations of the theory. In contrast to DFT a mathematically
rigorous formulation of TDDFT is missing. Here we will take a first
step towards this goal. We will give conditions for the existence of
a solution to the Sturm-Liouville boundary value problems at hand.
We will introduce the set of external potentials and time-dependent
densities under consideration. We will prove that all orders of the
Taylor expansions of the effective potentials exist, if the initial
configurations and the different two-particle interactions as well
as the external potential of the interacting system are spatially
infinitely differentiable. Given that these conditions hold, we will
show the set of $v$-representable time-dependent densities being purely
determined by the initial one-particle density and its first
derivative in time at $t=t_{0}$.
\\
\\
Section 2 summarizes the basic theorems of TDDFT so far. Section 3
outlines fundamental properties of the potentials and investigates
the existence of general solutions of the corresponding
Sturm-Liouville problems. In section 4 we introduce a certain set of
potentials and $v$-representable densities and prove the existence
of the solutions of all considered Sturm-Liouville
boundary value problems, and hence all orders of the Taylor
expansion of the effective potentials, under certain restrictions.
Assuming these conditions we can evidence properties of the set of
$v$-representable densities. Finally we conclude in section 5.

\section{Time-dependent density functional theory}

We consider a general many-body Hamiltonian in atomic units of the
form
\begin{eqnarray}
\label{Hamiltonian} \hat{H}(t) =  \hat{T} + \hat{V}_{\mathrm{int}} +
\hat{V}([v];t).
\end{eqnarray}
The operator $\hat{T} = \sum_{\sigma} \int  d^3 r
\hat{\psi}^{\dagger}_{\sigma}(r) \left(- \frac{1}{2}
\nabla^{2}\right) \hat{\psi}_{\sigma}(r)$ is the kinetic term,
$\hat{V}_{\mathrm{int}}=\frac{1}{2}\sum_{\sigma,\sigma'} \int \int
d^3 r d^3 r' v_{\mathrm{int}}(|r-r'|)
\hat{\psi}_{\sigma}^{\dagger}(r) \hat{\psi}_{\sigma'}^{\dagger}(r'
)\hat{\psi}_{\sigma'}(r') \hat{\psi}_{\sigma}(r)$ is the interaction
and $\hat{V}([v];t) = \sum_{\sigma} \int d^3r \; v(r,t) \;
\hat{\psi}^{\dagger}_{\sigma}(r) \hat{\psi}_{\sigma}(r)$ is the
external potential, where $\hat{\psi}^{\dagger}_{\sigma}(r)$ and
$\hat{\psi}_{\sigma}(r)$ are the creation and annihilation operator
with spin $\sigma$ \cite{Fetter}. The interaction potential
$v_{\mathrm{int}}(|r-r'|)$ is arbitrary but will usually be chosen
to be equal to the Coulomb interaction. The external one-particle
potential $v(r,t)$ typically consists of a static part, e.g. the
attractive Coulomb potential of a fixed nucleus, and a
time-dependent part, e.g. a laser pulse in dipole approximation and
length gauge. The time-dependent one-particle density is defined by
the expectation value of
\begin{eqnarray}
 \hat{n}(r) := \sum_{\sigma} \hat{\psi}^{\dagger}_{\sigma}(r) \hat{\psi}_{\sigma}(r)
\end{eqnarray}
with the time-dependent density matrix $\hat{\rho}(t)$, i.e.
\begin{eqnarray}
 n(r,t) = \left< \hat{n}(r) \right> = \tr \left( \hat{\rho}(t) \; \hat{n}(r) \right).
\end{eqnarray}
With the current-density operator
\begin{eqnarray}
 \hat{\jmath}(r) := \frac{1}{2\,\imagi} \sum_{\sigma} \left\{ \hat{\psi}_{\sigma}^{\dagger}(r) \nabla \hat{\psi}_{\sigma}(r) - \left[ \nabla \hat{\psi}_{\sigma}^{\dagger}(r)\right] \hat{\psi}_{\sigma}(r) \right\}
\end{eqnarray}
it is straightforward to find the usual continuity equation via
application of the Heisenberg equation
\begin{eqnarray}
\label{Continuity}
\partial_{t} n(r,t) & = & - \imagi \left< [\hat{n}(r), \hat{H}(t)]_{-} \right> = -\nabla \cdot j(r,t),
\end{eqnarray}
where $\partial_{t} \equiv \partial/ \partial t$ and
$[\cdot,\cdot]_{-}$ is the usual commutator. The time-derivative of
the current-density leads to the local force balance equation
\cite{TDDFT}
\begin{eqnarray}
\label{LocalForceBalance}
 \partial_{t} j_{\nu}(r,t) = - n(r,t) \partial_{\nu} v(r,t) - \partial_{\mu} \left< \hat{T}_{\mu \nu}(r) \right> -\left< \hat{W}_{ \nu}(r) \right>
\end{eqnarray}
where we made use of the Einstein summation convention, i.e. summing
over multiple indices, the momentum-stress-tensor
\begin{eqnarray}
 \hat{T}_{\mu \nu}(r) &:=& \frac{1}{2} \sum_{\sigma} \Bigg\{ \left(\partial_{\mu}  \hat{\psi}^{\dagger}_{\sigma}(r) \right) \partial_{\nu} \hat{\psi}_{\sigma}(r) + \left(\partial_{\nu}  \hat{\psi}^{\dagger}_{\sigma}(r) \right) \partial_{\mu} \hat{\psi}_{\sigma}(r)
\nonumber\\
&& -\frac{1}{2} \partial_{\mu} \partial_{\nu}
\left(\hat{\psi}^{\dagger}_{\sigma}(r)\hat{\psi}_{\sigma}(r)\right)
\Bigg\},
\end{eqnarray}
and the divergence of the interaction-stress-tensor \cite{Tokatly}
\begin{eqnarray}
 \hat{W}_{\nu} (r) := \sum_{\sigma, \sigma'} \int d^{3} r' \; \left(\partial_{\nu} v_{\mathrm{int}}(|r- r'|) \right) \hat{\psi}^{\dagger}_{\sigma}(r) \hat{\psi}^{\dagger}_{\sigma'}(r')\hat{\psi}_{\sigma'}(r')  \hat{\psi}_{\sigma}(r).
\end{eqnarray}
Now we will shortly sketch the idea underlying the Runge-Gross
proof. Assume two external potentials $v(r,t)$ and $v'(r,t)$, both
Taylor expandable about the initial time $t=t_{0}$, which differ by
more than a merely time-dependent function $c(t)$, i.e. $v(r,t) -
v'(r,t) \neq c(t)$. If we evolve an initial configuration
$\hat{\rho}(t_{0}) = \hat{\rho}_{0}$ of a finite multi-particle
system in time with the two different external potentials, we can
investigate the difference of the time-derivatives of the
current-densities at time $t=t_{0}$ via application of equation
(\ref{LocalForceBalance}). If the time-derivatives of the
current-densities differ for some order, then the corresponding
densities will be different after an infinitesimal time-step. This
leads to the Runge-Gross theorem \cite{Runge_Gross}:
\begin{theorem}
For every single-particle potential $v(r,t)$ which can be expanded
into a Taylor series with respect to the time coordinate around
$t=t_{0}$, a map $G:v(r,t) \mapsto n(r,t)$ is defined by solving the
time-dependent Schr\"odinger equations with fixed initial
configuration $\hat{\rho}(t_{0})=\hat{\rho}_{0}$ and calculating the
corresponding density $n(r,t)$. This map can be inverted up to an
additive, merely time-dependent function in the potential.
\end{theorem}
Now let us fix the additive merely time-dependent function in the
potential to be equal to zero by the boundary condition $v(r,t)
\rightarrow 0$ for $|r| \rightarrow \infty$. Hence we have an
invertible mapping $G$ which obviously depends on the initial
configuration $\hat{\rho}_{0}$. The proof is not restricted to
Coulombic interactions and can be applied to any reasonable
interaction. The domain and the range of this mapping is not further
investigated in the original paper \cite{Runge_Gross}. With the
mapping $F: v(r,t) \mapsto \hat{\rho}(t)$ defined by the solutions
of the associated Schr\"odinger equations we further find via $F
\circ G^{-1}: n(r,t) \mapsto \hat{\rho}(t)$ that every expectation
value of an operator $\hat{O}$, i.e. $O = \tr [\hat{O}
\hat{\rho}(t)]$, is uniquely determined by the density alone.
Although this theorem holds also for noninteracting systems, it is
not clear that every density subject to the Runge-Gross theorem in
an interacting system can be reproduced by some effective Kohn-Sham
potential in a noninteracting system. In other words, it is not
known if the interacting $v$-representable density is also
noninteracting $v$-representable. Initial attempts to construct such
connections in the original paper \cite{Runge_Gross} led to the
symmetry-causality paradox \cite{Gross95}.
\\
In order to overcome these problems the extended Runge-Gross theorem
was introduced in \cite{vanLeeuwen99}. A sketch of the proof reads
as follows:
\\
We apply the continuity equation (\ref{Continuity}) to the local
force balance equation (\ref{LocalForceBalance}), leading to
\begin{eqnarray}
\label{Continuity_2} \frac{\partial^{2}}{\partial t^2} n(r,t) =
\nabla \cdot \left[ n(r,t) \nabla v(r,t) \right] + \left<
\underbrace{\partial_{\nu} \left( \partial_{\mu} \hat{T}_{\mu
\nu}(r) + \hat{W}_{ \nu}(r) \right)}_{=:\hat{q}(r)} \right>.
\end{eqnarray}
If we now assume the external potential $v(r,t)$ as well as the
density $n(r,t)$ to be analytic about $t=t_{0}$ the
different orders of the Taylor expansion are connected via equation
(\ref{Continuity_2}) leading to \cite{TDDFT}
\begin{eqnarray}
\label{ConstDensity} n^{(k+2)}(r) = q^{(k)}(r) + \sum_{l=0}^{k}
\left( \begin{array}{c} k \\ l \end{array} \right) \nabla \cdot
\left[ n^{(k-l)}(r) \nabla v^{(l)}(r) \right],
\end{eqnarray}
for $k>1$ where we used $n^{(k)}(r) = \partial_{t}^k n(r,t)
|_{t=t_0}$ and $q^{(k)}(r)$ is defined by applying the Heisenberg
equation $k$ times to $\hat{q}(r)$ at time $t=t_{0}$. Hence,
$q^{(k)}(r)$ contains terms $v^{(l)}(r)$ up to order $l = k-1$. For
a second system with Hamiltonian
\begin{eqnarray}
 \hat{H}'(t)=\hat{T} + \hat{V}_{\mathrm{int}}' + \hat{V}'([v'];t)
\end{eqnarray}
and initial state $\hat{\rho}'(t_{0}) = \hat{\rho}'_{0}$ and
accordingly redefined operator $\hat{q}'(r)$ we can rederive
equation (\ref{ConstDensity}) for the primed system. Given the
time-dependent density $n(r,t)$ of the unprimed system we can define
the potential $v'(r,t)$ leading to the same density in the primed
system in terms of its Taylor expansion in form of a Sturm-Liouville
problem
\begin{eqnarray}
\label{v^k_def}
 \nabla \cdot \left[ n^{(0)}(r) \nabla v'^{(k)}(r)   \right] &=& n^{(k+2)}(r) - q'^{(k)}(r)
\nonumber\\
&&- \sum_{l=0}^{k-1} \left( \begin{array}{c} k \\ l \end{array}
\right) \nabla \cdot \left[ n^{(k-l)}(r) \nabla v'^{(l)}(r) \right].
\end{eqnarray}
With this we can state the extended Runge-Gross theorem \cite{vanLeeuwen99}:
\begin{theorem}
For a Hamiltonian $\hat{H}(t)$ with an analytic potential $v(r,t)$
about $t=t_{0}$ we assume the density $n(r,t)$ generated via
propagation of the initial configuration $\hat{\rho}_{0}$ to be
analytic about $t=t_{0}$ as well. For a second system $\hat{H}'(t)$
with initial configuration $\hat{\rho}_{0}$ subject to the
conditions
\numparts
\begin{eqnarray}
\label{IniCondRG1} n(r, t_0)= n^{(0)}(r)&=&n'(r, t_0),
\\
\label{IniCondRG2} \tr \left(\hat{\rho}_{0} \; \nabla \cdot
\hat{\jmath}(r) \right)&=&n^{(1)}(r) = \tr \left(\hat{\rho}'_0 \;
\nabla \cdot \hat{\jmath}(r) \right),
\end{eqnarray}
\endnumparts and an interaction $\hat{V}_{\mathrm{int}}'$ assumed
such that its expectation value and its derivatives are finite, the
analytic potential $v'(r,t)$ leading to the same density $n(r,t)$ is
uniquely defined up to a purely time-dependent function.
\end{theorem}
Again we can fix the purely time-dependent function of the
potentials to be equal to zero by choosing the boundary condition
$v'(r,t) \rightarrow 0$ for $|r| \rightarrow \infty$. The existence
of this potential $v'(r,t)$ is investigated in \cite{vanLeeuwen01}
by minimization of a corresponding functional. However, a rigorous
proof of existence was not given.
\\
Note, in the above schematically depicted proof of the extended
Runge-Gross theorem one made use of the knowledge of the density
$n(r,t)$. However, we want a theory \textit{predicting} the density. That
TDDFT is a predictive theory can be seen if we use equation
(\ref{ConstDensity}) for the primed and the unprimed system and
assume both to yield the same density $n(r,t)$. With the definition
\begin{eqnarray}
 v'(r,t) = v(r,t) + v_{\Delta}(r,t)
\end{eqnarray}
we thus infer the Sturm-Liouville problem
\begin{eqnarray}
\label{HxcPotential}
 \nabla \cdot &&  \!\!\left[ n^{(0)}(r) \nabla v_{\Delta}^{(k)}([n]; r) \right]
 = \zeta^{(k)}(r) :=
\nonumber\\
 &&q^{(k)}(r) - q'^{(k)}(r) - \sum_{l=0}^{k-1} \left( \begin{array}{c} k \\ l \end{array} \right) \nabla \cdot \left[ n^{(k-l)}(r) \nabla v_{\Delta}^{(l)}([n]; r)   \right].
\end{eqnarray}
Via equation (\ref{HxcPotential}) one can generate $v'^{(k)}(r)$
which then can be used in equation (\ref{ConstDensity}) to find the
corresponding higher terms of the Taylor expansion of $n(r,t)$. In
return, the higher terms of the density Taylor expansion again
determine the next term in the Taylor expansion of the potential.
Hence, only the initial configurations and the external potential of
the unprimed system are needed to generate the expansion of the
density about $t=t_{0}$. However, the special form of $q^{(k)}(r)$
defined by successive application of the Heisenberg equation led to
the question wether a simultaneous solution of the unprimed system
is required in principle \cite{Maitra, Schirmer}.

\section{Properties of the potentials and the Sturm-Liouville problem}

In order to define the sets of one-particle potentials and the
associated one-particle densities, i.e. the $v$-representable
densities, we start by looking at the properties of the
corresponding Hamiltonian $\hat{H}(t)$. We demand the Hamiltonian to
be self-adjoint for every time $t$ on the domain of the kinetic
energy operator $\mathrm{dom}(\hat{T})$. The free Hamiltonian
$\hat{H}_{0} = \hat{T} + \hat{V}_{\mathrm{int}}$ with
$\hat{V}_{\mathrm{int}}$ the Coulomb interaction can be shown to
fulfill this constraint \cite{Blanchard_Bruening} and any other
interaction $\hat{V}_{\mathrm{int}}'$ under consideration will be
assumed to do so as well. This condition is trivially fulfilled if
we choose $\hat{V}_{\mathrm{int}}' \equiv 0$. With the theory of
Kato perturbations \cite{Blanchard_Bruening} we find the simple
condition
\begin{eqnarray}
\label{SelfAdjCondition}
 v(t) \in L^{2}(\mathbb{R}^{3}) + L^{\infty}(\mathbb{R}^{3})
\end{eqnarray}
for every time $t$. Here $L^{2}$ and $L^{\infty}$ are the usual
Lebesgue quotient spaces with norm $\| . \|_{2}$ and $\| .
\|_{\infty}$, respectively. $L^{2}(\mathbb{R}^{3}) +
L^{\infty}(\mathbb{R}^{3})$ is a Banach space with the norm
\begin{eqnarray}
 \| v(t) \| = \inf \biggl\{  \|v_{1}(t)\|_{2} + \|v_{2}(t)\|_{\infty} \,\biggl|\,&& v_{1}(t) \in L^{2}(\mathbb{R}^{3}), \; v_{2}(t) \in L^{\infty}(\mathbb{R}^{3}),
\nonumber\\
&& \; v(t) = v_{1}(t) + v_{2}(t) \biggr\}.
\end{eqnarray}

The effective potentials $v'(r,t)$ leading to the same density as in
the Coulombic system exists in accordance to the extended
Runge-Gross proof if all orders of equation (\ref{v^k_def}) have an
existing solution. Accordingly, this is true if all orders of
equation (\ref{HxcPotential}) have a solution with $v'(r,t) =
v(r,t) + v_{\Delta}(r,t)$. A way to investigate the existence of
such solutions to such Sturm-Liouville boundary value problems is
within the following framework.\\

Let us first introduce the bounded open domain $\Omega \subset
\mathbb{R}^{3}$ with piecewise $\mathcal{C}^1$ boundary $\partial
\Omega$. For the real Hilbert space $L^{2}(\Omega)$ with scalar product denoted as $\langle
\cdot,\cdot \rangle_2$ and norm $\|\cdot\|_2$ we define the Sobolev
space \cite{Blanchard_Bruening}
\begin{equation}
 W^{1,2}(\Omega) = \left\{ u \in L^{2}(\Omega) \biggl|\; \partial_{j} u \in L^{2}(\Omega),\; j = 1,...,3 \right\}
\end{equation}
with norm $\|u\|_{1,2}^2 = \|u\|_2^2 + \|\nabla u\|_2^2$. Here the
partial derivatives $\partial_{j} u$ are understood in the
distributional sense. Further the real Hilbert space $H^{1}_{0}(\Omega)$
is defined to be the closure in $W^{1,2}(\Omega)$ of the infinitely
differentiable functions compactly supported in $\Omega$. Hence,
functions $u \in H^{1}_{0}(\Omega)$ fulfill the boundary condition
$u = 0 $ on $\partial \Omega$ naturally. The previously considered
Sturm-Liouville problem (\ref{HxcPotential}) in shorthand notation
reads as
\begin{equation}
\label{SturmLiouville}
 \nabla \cdot \left( n \nabla v \right) = \zeta
\end{equation}
and defines a bilinear form $Q$ on $H^{1}_{0}(\Omega)$ by
\begin{equation}
 Q(u,v) = \langle u, -\nabla \cdot \left( n \nabla v \right)
 \rangle_2 = \langle \nabla u, n \nabla v \rangle_2
\end{equation}
where we used integration by parts and that functions in
$H^{1}_{0}(\Omega)$ vanish at the border. We take advantage of the
fact that (\ref{SturmLiouville}) has a (weak) solution $v \in
H^{1}_{0}(\Omega)$ if, and only if, $Q(u, v) = -\langle u,\zeta
\rangle_2$ for all $u \in H^{1}_{0}(\Omega)$. This immediately leads
us to the necessity $\zeta \in L^{2}(\Omega)$. Now the answer to the
question of solvability is at hand with the Theorem of Lax-Milgram.
\cite{Blanchard_Bruening2}

\begin{theorem}[Lax-Milgram]
\label{LaxMilgram} Let $Q$ be a coercive continuous bilinear form on
a Hilbert space $\mathcal{H}$. Then for every continuous linear
functional $f$ on $\mathcal{H}$, there exists a unique $u_f \in
\mathcal{H}$ such that
\begin{equation}
 Q(u,u_f) = f(u)
\end{equation}
holds for all $u \in \mathcal{H}$.
\end{theorem}

A bilinear form $Q$ is said to be coercive if there exists a
constant $c > 0$ such that $Q(u,u) \geq c \|u\|^2$ for all $u \in
\mathcal{H}$. In our case this can be established by means of the
Poincar\'{e} inequality
\begin{equation}
 \|u\|_2 \leq \lambda \|\nabla u\|_2,\quad \forall u \in H^{1}_{0}(\Omega)
\end{equation}
where $0 < \lambda = \lambda(\Omega) < \infty$. As an additional
assumption we add that $n$ is bounded by a constant $m > 0$ almost
everywhere on $\Omega$ from below. Then \numparts
\begin{eqnarray}
 Q(u,u) = \langle \nabla u, n \nabla u \rangle_2 \geq m \|\nabla
 u\|_2^2 \\
 \lambda^2 Q(u,u) \geq \lambda^2 m \|\nabla u\|_2^2 \geq m \|u\|_2^2
\end{eqnarray}
\endnumparts

Combination of these results yields
\begin{equation}
 Q(u,u) \geq \frac{m}{1+\lambda^2} \left( \|u\|_2^2 + \|\nabla u\|_2^2
 \right) = \frac{m}{1+\lambda^2} \| u \|_{1,2}^2.
\end{equation}
We thus have established the coercivity of $Q$. As for the
continuity we add another assumption on $n$ that is boundedness from
above by a constant $M > 0$ almost everywhere on $\Omega$.
\begin{eqnarray}
 | Q(u,v) &-& Q(u_0,v) | = | Q(u-u_0,v) | =
 |\langle \nabla (u-u_0), n \nabla u \rangle_2 | \nonumber\\
 & \leq & M \|\nabla (u-u_0)\|_2 \cdot \|\nabla v\|_2 \leq
 M \|u-u_0\|_{1,2} \cdot \|v\|_{1,2} < \infty
\end{eqnarray}

These restrictions on $n$ also imply that the differential operator
defined by the left hand side of (\ref{SturmLiouville}) is elliptic.
If $n$ can be assumed to be continuous on the closed domain
$\bar{\Omega}$, then $n$ also attains its extremal values on
$\bar{\Omega}$ and the restrictions reduce to the form
\begin{equation}
 0 < n < \infty \quad \mbox{on}\; \bar{\Omega}.
\end{equation}

Now everything left to show for the application of Theorem
\ref{LaxMilgram} is that the right hand side $-\langle \cdot,\zeta
\rangle_2$ is indeed a continuous linear functional on the real Hilbert space
$H^{1}_{0}(\Omega)$. This is easily established by considering for
arbitrary $u, u_0 \in H^{1}_{0}(\Omega)$
\begin{eqnarray}
\label{ZetaContinuous}
 | \langle u,\zeta \rangle_2 &-& \langle u_0,\zeta \rangle_2
 | = | \langle u-u_0,\zeta \rangle_2 | \nonumber\\
 &\leq& \|u-u_0\|_2 \cdot
 \|\zeta\|_2 \leq \|u-u_0\|_{1,2} \cdot \|\zeta\|_2 < \infty.
\end{eqnarray}

We subsume our results for the solvability of the Sturm-Liouville
problem (\ref{SturmLiouville}) in the following corollary.

\begin{corollary}
\label{WeakCorollary}
 Consider the Sturm-Liouville problem $\nabla \cdot \left( n \nabla v \right) =
 \zeta$ on the bounded open domain $\Omega \subset \mathbb{R}^{3}$ with piecewise
 $\mathcal{C}^1$ boundary. Let $\zeta \in
 L^{2}(\Omega)$ and $n : \Omega \rightarrow \mathbb{R}$
 be almost everywhere bounded by $0 < m \leq n \leq M < \infty$. Then there exists a
 unique solution $v \in H^{1}_{0}(\Omega)$.
\end{corollary}

Looking back to the original formulation of the Sturm-Liouville
problem (\ref{HxcPotential}) we note that if $n^{(0)}$ fulfills the
prerequisites of $n$ in Corollary \ref{WeakCorollary} and the right
hand side is indeed in $L^{2}(\Omega)$ then all orders of the
potential $v_{\Delta}^{(k)}$ are uniquely defined in
$H^{1}_{0}(\Omega)$. We now want to examine which constraints on the
higher orders $n^{(k)}$ are sufficient for $\zeta^{(k)} \in
L^{2}(\Omega)$. For this we assume $q^{(k)}$ and $q'^{(k)}$ already
in $L^{2}(\Omega)$ for all $k$ and thus we get a unique
$v_{\Delta}^{(0)} \in H^{1}_{0}(\Omega)$ trivially by Corollary
\ref{WeakCorollary}. For $k>1$ we apply inductive reasoning: Let us
assume $v_{\Delta}^{(l)} \in H^{1}_{0}(\Omega)$ be given uniquely
for $l < k$. Then (\ref{HxcPotential}) in a shorter notation reads
as
\begin{equation}
 \nabla \cdot \left( n^{(0)} \nabla v_{\Delta}^{(k)} \right)
 = q^{(k)} - q'^{(k)} - \sum_{l=0}^{k-1} \left( \begin{array}{c} k \\ l \end{array} \right) \nabla \cdot \left( n^{(k-l)} \nabla v_{\Delta}^{(l)} \right).
\end{equation}

The (elliptic) differential operator on the left defines the same
coercive continuous bilinear form on $H^{1}_{0}(\Omega)$ as before,
thus everything to show for an application of Theorem
\ref{LaxMilgram} is that the right hand side yields a continuous
linear functional. The sum of such functionals is again linear
continuous therefore we can examine all terms separately. For
$q^{(k)}$ and $q'^{(k)}$ the same reasoning as in
(\ref{ZetaContinuous}) applies. Finally with $u, u_0 \in
H^{1}_{0}(\Omega)$ continuity of the separate terms of the sum is
established by
\begin{eqnarray}
\left| \left\langle u, \nabla \cdot \left( n^{(k-l)} \nabla
v_{\Delta}^{(l)} \right) \right\rangle_2 - \left\langle u_0, \nabla
\cdot \left( n^{(k-l)} \nabla v_{\Delta}^{(l)} \right)
\right\rangle_2 \right| \nonumber\\
\quad= \left| \left\langle \nabla (u-u_0), n^{(k-l)} \nabla
v_{\Delta}^{(l)} \right\rangle_2 \right| \leq \| \nabla (u-u_0) \|_2
\cdot \| n^{(k-l)} \nabla v_{\Delta}^{(l)} \|_2 \nonumber\\
\quad\leq \| u-u_0 \|_{1,2} \cdot \| n^{(k-l)} \nabla
v_{\Delta}^{(l)} \|_2 < \infty \quad \mbox{if} \quad n^{(k-l)}
\nabla v_{\Delta}^{(l)} \in L^{2}(\Omega).
\end{eqnarray}

By induction we know that $v_{\Delta}^{(l)} \in H^{1}_{0}(\Omega)$
and thus $\nabla v_{\Delta}^{(l)} \in L^{2}(\Omega)$ so $n^{(k)}$
bounded almost everywhere by some $M_k > 0$ for all $k$ is a
sufficient condition for Theorem \ref{LaxMilgram} to be applied. We
subsume this in a second corollary.

\begin{corollary}
\label{WeakCorollary2}
 Consider the system of Sturm-Liouville problems (\ref{HxcPotential}) on the bounded open domain $\Omega \subset \mathbb{R}^{3}$ with piecewise
 $\mathcal{C}^1$ boundary. Let $q^{(k)}, q'^{(k)} \in
 L^{2}(\Omega)$ and $n^{(k)} : \Omega \rightarrow \mathbb{R}$
 be almost everywhere bounded by $n^{(k)} \leq M_k$ with $M_k > 0$ for all $k$ and additionally $n^{(0)} \geq m > 0$ almost everywhere. Then there exists a
 unique sequence of solutions $v_{\Delta}^{(k)} \in H^{1}_{0}(\Omega)$.
\end{corollary}

If we turn to the problem of a classical solution we refer the
reader to the Weyl Lemma as given in \cite{Hellwig, Hellwig2} in
various forms. (In a more general formulation it can be found as the
``fundamental theorem on weak solutions" in \cite{Maurin}.) There
the operator $\hat{K} v = \nabla \cdot (n \nabla v)$ has domain
\begin{eqnarray}
 \mathrm{dom}(\hat{K}) = \left\{ u \biggl| u \in \mathcal{C}^{1}(\bar{\Omega}), \; u \in \mathcal{C}^{2}(\Omega) ,\; \hat{K} u \in L^{2}(\Omega) ; \; u=0 \;\mbox{on}\; \partial \Omega  \right\}.
\end{eqnarray}

\begin{theorem}
\label{Existence_StuLiou} For $n > 0$ on $\bar{\Omega}$ and $n \in
\mathcal{C}^{3}(\bar{\Omega})$ and the boundary condition $v = 0$ on
$\partial \Omega$, the equation
\begin{eqnarray}
 \nabla \cdot (n \nabla v) = \zeta
\end{eqnarray}
has a classical solution $v \in \mathrm{dom}(\hat{K})$ if $\zeta \in
\mathcal{C}^{1}(\bar{\Omega})$ or $\zeta$ is H\"older continuous,
i.e. $|\zeta(r)-\zeta(r')| \leq h |r-r'|^{\alpha}$ for all $r,\; r'
\in \bar{\Omega}$ with $h$ and $0 < \alpha < 1$ independent of $r,\;
r'$.
\end{theorem}
Wether $\zeta^{(k)}$ in our actual problem (\ref{HxcPotential})
fulfills one of the conditions for a weak or classical solution,
respectively, depends on the properties of the initial
configurations and on the interactions under consideration. The
terms $q^{(k)}$ and $q'^{(k)}$ implicate already for $k=0$ spatial
partial derivatives of order 4 and spatial partial derivatives of
the involved interaction potential of order 3. Hence, to have a well
defined Sturm-Liouville problem, the wave-functions of the initial
configurations and the interaction potentials have to fulfill
certain restrictions with respect to their spatial behavior.

\section{Sets of potentials and $v$-representable densities}

We will introduce the set of external potentials for the extended
Runge-Gross theorem in accordance to the classical Sturm-Liouville
theory. Therefore, the defined sets will only be subsets of the
actual sets of $v$-representable densities and potentials connected
via the extended Runge-Gross theorem. Further, we will restrict our
considerations on the above introduced domain $\Omega \subset
\mathbb{R}^{3}$. We assume the boundary to be far away from the
center of the system such that it will not influence the dynamics.
One has to be careful at this point as we will assume the initial
one-particle density to be nonzero on $\bar{\Omega}$, hence also on
the boundary $\partial \Omega$. This is different to the usual
notion of a physical system restricted to a finite region, where one
assumes an infinite boundary potential to restrict the wave-function
to this domain. In this case the wave-function and thus the
one-particle potential will be zero at the boundary.
\\
For a free Hamiltonian $\hat{H}_{0}=\hat{T}+\hat{V}_{\mathrm{int}}$
assumed self-adjoint and an initial configuration $\hat{\rho}_{0}$
at time $t=t_{0}$ we have
\begin{eqnarray}
 \mathcal{V}(\hat{\rho}_{0},&& \hat{V}_{\mathrm{int}}) := \biggl\{ v \,\biggl|\, v \;\mbox{analytic about}\; t=t_{0}, v(t) \in
 \mathrm{dom}(\hat{K}), \nonumber
\\
&&  v \;\mbox{real};\; n[v] \;\mbox{analytic about}\;
t=t_{0} \;\mbox{for}\; \hat{\rho}_{0} \;\mbox{and}\;
\hat{V}_{\mathrm{int}} \biggr\}.
\end{eqnarray}
Here $n[v]$ is the time-dependent density, defined via the
propagation of the initial configuration $\hat{\rho}_{0}$ with the
Hamiltonian $\hat{H}(t)=\hat{T}+\hat{V}_{\mathrm{int}} +
\hat{V}([v];t)$. It is straightforward to proof self-adjointness of
this Hamiltonian by application of the Kato perturbation theory
\cite{Blanchard_Bruening} as one can use $v(t) \in
L^{\infty}(\Omega)$. Further we define the set of $v$-representable
variations by
\begin{eqnarray}
 \delta \mathcal{N}(\hat{\rho}_{0}, \hat{V}_{\mathrm{int}}):= \biggl\{ \delta n \,\biggl|\,&& \delta n(r,t) = \sum_{k=2}^{\infty} \frac{1}{k!}  n^{(k)}([v];r) (t-t_{0})^{k} \;\mbox{for}\; \hat{\rho}_{0}
 \;\mbox{and}\; \hat{V}_{\mathrm{int}}, \nonumber
\\
&& v \in \mathcal{V}(\hat{\rho}_{0}, \hat{V}_{\mathrm{int}})
\biggr\}.
\end{eqnarray}
The set of $v$-representable densities is an affine set
\begin{eqnarray}
\mathcal{N}(\hat{\rho}_{0}, \hat{V}_{\mathrm{int}}) : = n^{(0)} (r)
+ n^{(1)}(r,t) + \delta \mathcal{N}(\hat{\rho}_{0},
\hat{V}_{\mathrm{int}}),
\end{eqnarray}
where $n^{(0)}(r) = n(r,t_{0})$ is the initial density and
$n^{(1)}(r,t) = \mathrm{tr}(\hat{\rho}_{0} \nabla \cdot
\hat{\jmath}(r)) (t-t_{0})$ in accordance to (\ref{IniCondRG1}) and
(\ref{IniCondRG2}), respectively. For these sets we then have in
accordance to the Runge-Gross theorem an invertible mapping
\begin{eqnarray}
 v_{\hat{\rho}_{0}}: \mathcal{N}(\hat{\rho}_{0}, \hat{V}_{\mathrm{int}}) &\rightarrow& \mathcal{V}(\hat{\rho}_{0}, \hat{V}_{\mathrm{int}})
\\
 \qquad \quad \; \; \; n(r,t)&\mapsto&v_{\hat{\rho}_{0}}([n];
 r,t) \nonumber
\end{eqnarray}
connecting the $v$-representable one-particle densities with the
external potentials. Nevertheless, if we now define a second mapping
for a different initial configuration $\hat{\rho}_{0}'$ subject to
the conditions (\ref{IniCondRG1}) and (\ref{IniCondRG2}), and a
different interaction $\hat{V}'_{\mathrm{int}}$, we do not know if
$n$ is simultaneously element in $\mathcal{N}(\hat{\rho}_{0},
\hat{V}_{\mathrm{int}})$ and $\mathcal{N}(\hat{\rho}'_{0},
\hat{V}'_{\mathrm{int}})$. This, however, is of fundamental
importance if we want a rigorous formulation of the time-dependent
Kohn-Sham scheme.
\\
\\
To achieve this goal we introduce further restrictions. We will
assume smooth interactions and initial configurations in what
follows. This excludes the usual Coulombic interaction as it is not
infinitely differentiable at the origin. However, one may
regularize the Coulombic interaction by a so-called soft-core
interaction, i.e. by replacing $|r| \rightarrow \sqrt{r^{2} +
\epsilon} $ and $\epsilon > 0$.
\\
Due to equation (\ref{ConstDensity}) we have a direct connection between $v$-representable densities and potentials. Hence, we can formulate the following lemma
\begin{lemma}
\label{DensityExist} Let $\hat{\rho}_{0}$ be chosen such that all
its wavefunctions are in $\mathcal{C}^{\infty}(\bar{\Omega}^{N})$,
$v\; \mathrm{Taylor \; expandable \; about} \; t=t_{0}, \; v^{(k)}
\in \mathcal{C}^{\infty}(\bar{\Omega}) \; \forall \; k,$ and
$v_{\mathrm{int}}(|r-r'|)$ infinitely differentiable. Then
$n^{(k)}(r) \in \mathcal{C}^{\infty}(\bar{\Omega})$ for all $k$.
\end{lemma}
{\bf Proof.} We will use equation (\ref{ConstDensity}). Obviously we
have $n^{(0)}(r)$ and $n^{(1)}(r)$ in
$\mathcal{C}^{\infty}(\bar{\Omega})$. Thus $n^{(2)}(r)$ is in
$\mathcal{C}^{\infty}(\bar{\Omega})$ if $q^{(0)}(r)$ is infinitely
differentiable, where
\begin{eqnarray}
 q^{(0)}(r)= \tr \biggl[\hat{\rho}_{0} \; \hat{q}(r) \biggr].
\end{eqnarray}
$\hat{q} (r)$ consists of partial derivatives with respect to $r$
and of derivatives of $v_{\mathrm{int}}(|r-r'|)$. We have assumed
$v_{\mathrm{int}}(|r-r'|)$ infinitely differentiable. Hence, we have
$q^{(0)}(r) \in \mathcal{C}^{\infty}(\bar{\Omega})$. For
$n^{(3)}(r)$ we need to know $q^{(1)}(r)$. This is the commutator of
$\hat{q}(r)$ with $\hat{H}(t)$ at $t=t_{0}$. All functions in
$\hat{H}(t)$ are infinitely differentiable. Again the above
reasoning applies, and we find $q^{(1)}(r) \in
\mathcal{C}^{\infty}(\bar{\Omega})$. All higher terms are to be
found via successive application of the Heisenberg equation for
$\hat{q}(r)$ with $\hat{H}(t)$ at $t=t_{0}$. The only difference to
the above reasoning is the appearance of $v^{(k)}(r)$-terms, which
are again infinitely differentiable. Therefore one can successively
construct all $n^{(k)}(r) \in \mathcal{C}^{\infty}(\bar{\Omega})$.
$\Box$
\\
\\
Now we introduce the restricted set of smooth one-particle potentials
\begin{eqnarray}
 \mathcal{V}^{*}(\hat{\rho}_{0}, \hat{V}_{\mathrm{int}}) = && \biggl\{ v \,\biggl|\, v \in \mathcal{V}(\hat{\rho}_{0}, \hat{V}_{\mathrm{int}}), \; v(t) \in \mathcal{C}^{\infty}(\bar{\Omega})  \biggr\}.
\end{eqnarray}
and by lemma \ref{DensityExist} the corresponding smooth $v$-representable one-particle densities
\begin{eqnarray}
\mathcal{N}^{*}(\hat{\rho}_{0}, \hat{V}_{\mathrm{int}}) = \biggl\{ n
\,\biggl|\, && n(r,t) = \sum_{k=0}^{\infty} \frac{1}{k!}
n^{(k)}([v];r) (t-t_{0})^{k} \;\mbox{for}\;
\hat{\rho}_{0}\;\mbox{and}\; \hat{V}_{\mathrm{int}}, \nonumber
\\
&& v \in \mathcal{V}^{*}(\hat{\rho}_{0}, \hat{V}_{\mathrm{int}})
\biggr\}.
\end{eqnarray}
With this we can reformulate the extended Runge-Gross theorem as
follows.
\begin{theorem}
Let $\hat{\rho}_{0}$ and $\hat{V}_{\mathrm{int}}$ be infinitely
differentiable, $n(r,t) \in \mathcal{N}^{*}(\hat{\rho}_{0},
\hat{V}_{\mathrm{int}})$ and $v_{\hat{\rho}_{0}}([n]; r,t)=v(r,t)
\in \mathcal{V}^{*}(\hat{\rho}_{0}, \hat{V}_{\mathrm{int}})$ the
associated external potential. For a system with infinitely
differentiable interaction $\hat{V}'_{\mathrm{int}}$ and the initial
configuration $\hat{\rho}_{0}'$ consisting of infinitely
differentiable functions subject to the constraint
\begin{eqnarray}
n(r, t_0) = n^{(0)}(r) &=& n'(r, t_0) > 0,
\\
\tr \left(  \hat{\rho}_{0} \; \nabla \cdot \hat{\jmath}(r) \right)
&=& \tr \left(  \hat{\rho}'_0 \; \nabla \cdot \hat{\jmath}(r)
\right),
\end{eqnarray}
there exists a unique effective potential depending on both initial configurations
\begin{eqnarray}
 v_{\hat{\rho}_{0},\hat{\rho}_{0}'}([n]; r,t) = \sum_{k=0}^{\infty} \frac{1}{k!} \; v_{\Delta}^{(k)}(r) \;\; (t-t_{0})^{k},
\end{eqnarray}
where $v_{\Delta}^{(k)}(r)$ is defined via
\begin{eqnarray}
 \nabla \cdot &&\left[ n^{(0)}(r) \nabla v_{\Delta}^{(k)}(r) \right]
 = \nonumber
\\
 &&q^{(k)}(r) - q'^{(k)}(r) - \sum_{l=0}^{k-1} \left( \begin{array}{c} k \\ l \end{array} \right) \nabla \cdot \left[ n'^{(k-l)}(r) \nabla v'^{(k)}(r)   \right],
\end{eqnarray}
with $v' = \left( v + v_{\hat{\rho}_{0},\hat{\rho}_{0}'} \right) \in
\mathcal{V}^{*}(\hat{\rho}_{0}', \hat{V}_{\mathrm{int}}')$
generating the same density. It holds that
\begin{eqnarray}
  \mathcal{N}^{*}(\hat{\rho}_{0}, \hat{V}_{\mathrm{int}}) =  \mathcal{N}^{*}(\hat{\rho}_{0}', \hat{V}_{\mathrm{int}}')=\mathcal{N}^{*}(n^{(0)}, n^{(1)}).
\end{eqnarray}
\end{theorem}
{\bf Proof.} From the proof of lemma \ref{DensityExist} we know that
all $q^{(k)}$ and $q'^{(k)}$ are infinitely differentiable. As we
have assumed $n_{0}(r) > 0$ we can apply theorem
\ref{Existence_StuLiou} from which it is clear that
\begin{eqnarray}
 \nabla \cdot &&\left[ n^{(0)}(r) \nabla v'^{(k)}(r) \right] =
 \nonumber
\\
 &&q^{(k)}(r) - q'^{(k)}(r) - \sum_{l=0}^{k-1} \left( \begin{array}{c} k \\ l \end{array} \right) \nabla \cdot \left[ n'^{(k-l)}(r) \nabla v'^{(k)}(r)   \right]
\end{eqnarray}
has an existing solution for $k=0$ if the right hand side is
$\mathcal{C}^{1}(\bar{\Omega})$. Obviously $v'^{(0)}(r)$ exists due
to theorem \ref{Existence_StuLiou} and is infinitely differentiable.
In the next step we can use $v'^{(0)}(r)$ in the Sturm-Liouville
equation defining $v'^{(1)}(r)$. Again existence is guaranteed and
we have $v'^{(1)}(r) \in \mathcal{C}^{\infty}(\bar{\Omega})$
\cite{Agmon}. One can now successively construct
$v_{\hat{\rho}_{0},\hat{\rho}_{0}'}$. Then
$(v+v_{\hat{\rho}_{0},\hat{\rho}_{0}'})$ is given via its Taylor
series within its radius of convergence in accordance to the
extended Runge-Gross proof \cite{vanLeeuwen99}. This construction
holds for every $n \in \mathcal{N}^{*}(\hat{\rho}_{0},
\hat{V}_{\mathrm{int}})$, and we have $n \in
\mathcal{N}^{*}(\hat{\rho}_{0}', \hat{V}_{\mathrm{int}}')$ as well.
Hence, the set of $v$-representable densities does not depend on the
smooth interaction or on the smooth initial configuration. $\Box$
\\
\\
Here it became obvious why we restricted our considerations to
infinitely differentiable initial configurations and potentials.
With this assumptions we can guarantee the existence of all the
classical Sturm-Liouville boundary value problems on $\Omega$. For the general, i.e., weak case, we need to make sure that all $q^{(k)}, q'^{(k)}$ and hence $\zeta^{(k)}$ are in $L^{2}(\Omega)$, in order to proof the existence of a solution using corollary \ref{WeakCorollary2}.
\\
\\
The special case of a rigorous Kohn-Sham theorem is straightforward
as $\hat{V}_{\mathrm{int}} \equiv 0$ is of course infinitely
differentiable. One finds that for the above
restrictions all interacting-$v$-representable densities are
noninteracting-$v$-representable because $\mathcal{N}^{*}(\hat{\rho}_{0}, \hat{V}_{\mathrm{int}}) =  \mathcal{N}^{*}(\hat{\rho}_{0}', 0)$. Only a noninteracting initial configuration is needed. The condition of $n^{(0)}(r) > 0$ for the existence of the
effective potential may be relaxed if there exists some time $t_{1}$
in a sufficiently small neighbourhood of $t_0$ for which $n(r,t_{1})
> 0$. Then we could use $\hat{\rho}(t_{1})$ as new initial state and
prove existence at that time provided we also have the corresponding
$\hat{\rho}'(t_{1})$.

\section{Conclusion}

Under certain assumptions we can state sufficient constraints on the
one-particle density such that the existence of the effective
potential, possibly in the weak sense, is guaranteed. However, only
for classical solutions of the corresponding Sturm-Liouville
boundary value problems we can reformulate the extended Runge-Gross
theorem such that existence of the effective potentials is granted.
As long as we consider smooth initial states and smooth interactions
the Kohn-Sham system exactly reproduces the physical one-particle
density. In general, as pointed out in \cite{vanLeeuwen01}, it seems
safe to assume existence of the Kohn-Sham potential for physical
systems. Nevertheless, a rigorous proof of principle is of
importance for the foundations of the theory.

\section*{Acknowledgements}
This work was supported by the Deutsche Forschungsgemeinschaft.

\end{document}